\documentclass[aps,prb,twocolumn,preprintnumbers,superscriptaddress]{revtex4}
\usepackage{epsfig}

\newcommand{\Bra}[1]{{\left\langle{#1}\right|}}
\newcommand{\Ket}[1]{{\left|{#1}\right\rangle}}
\newcommand{\BraKet}[2]{{\left\langle{#1}\right|\left.\!{#2}\right\rangle}}
\newcommand{\BraOpKet}[3]{{\Bra{#1}{#2}\Ket{#3}}}

\preprint{PREPRINT}

\begin{document}
\title{Intersubband gain in a Bloch oscillator and Quantum cascade laser}
\author{H.~Willenberg}\email{harald.willenberg@unine.ch}\affiliation{Institut de Physique, Universit\'{e} de Neuch\^{a}tel,
Switzerland}\affiliation{Institut f\"ur technische Physik,
Universit\"at Erlangen, Germany}
\author{G.H.~D\"ohler}
\affiliation{Institut f\"ur technische Physik, Universit\"at
Erlangen, Germany}
\author{J.~Faist}\affiliation{Institut de Physique, Universit\'{e} de Neuch\^{a}tel,
Switzerland}

\date{\today}
\begin{abstract}
The link between the inversion gain of quantum cascade structures
and the Bloch gain in periodic superlattices is presented. The
proposed theoretical model based on the density matrix formalism
is able to treat the gain mechanism of the Bloch oscillator and
Quantum cascade laser on the same footing by taking into account
in-plane momentum relaxation. The model predicts a dispersive
contribution in addition to the (usual)
population-inversion-dependent intersubband gain in quantum
cascade structures and -- in the absence of inversion -- provides
the quantum mechanical description for the dispersive gain in
superlattices. It corroborates the predictions of the
semi-classical miniband picture, according to which gain is
predicted for photon energies lower than the Bloch oscillation
frequency, whereas net absorption is expected at higher photon
energies, as a description which is valid in the high-temperature
limit. A red-shift of the amplified emission with respect to the
resonant transition energy results from the dispersive gain
contribution in any intersubband transition, for which the
population inversion is small.
\end{abstract}

\maketitle

\section{Introduction}
Soon after the original proposal of semiconductor superlattices
\cite{Esaki:IJRD:70:61}, two apparently quite different schemes to
obtain optical gain in such novel systems were put forward. In a
seminal work, Kazarinov and Suris pointed out how to achieve a
population inversion, a key ingredient to obtain light
amplification, between electronic subbands in a strongly biased
superlattice \cite{Kazarinov:SPS:71:707}. On the other hand, based
on semi-classical arguments, Ktitorov \emph{et al.}
\cite{Ktitorov:FTT:71:2230} and later Ignatov \emph{et al.}
\cite{Ignatov:PSSB:76:327} predicted optical gain due to Bloch
oscillations within a miniband -- despite a missing population
inversion.

Two decades later, the demonstration of the quantum cascade laser
\cite{Faist:Sci:94:553} affirmed the first proposal. The
(conduction) band structure in each period is carefully designed
to allow for injecting electrons into an upper subband state, with
a long non-radiative lifetime, and to enable a fast extraction of
electrons from an accordingly tailored lower state. As a
consequence, population inversion is achieved. Also, by a suitable
design, the structure is electrically stable at threshold. By now,
the quantum cascade laser technology covers a wide range of the
electro-magnetic spectrum. Recently, a room-temperature
continuous-wave laser \cite{Beck:Sci:02:301} emitting at a
wavelength of $9\,\rm\mu m$ has been demonstrated and stimulated
emission in the terahertz regime at about $66\,\rm\mu m$
\cite{Koehler:Nat:02:156,Rochat:APL:02:submitted} has been
observed.

In contrast to this, the feasibility of the Bloch oscillator,
emitting electro-magnetic radiation, tunable by the external
electric dc field, is still under question. Besides the task to
stabilize the electric field domains in a biased superlattice at
the point of operation, the description of the gain mechanism is,
so far, based on semi-classical models only. In a naive picture,
electro-magnetic radiation of photon energy
$\hbar\omega\approx\hbar\omega_b=eFd$ is expected, corresponding
to the frequency of Bloch oscillations $\omega_b$ that linearly
depends on the applied dc field $F$. For, \emph{e.g.} a
superlattice period $d$ of some nanometers and fields of several
tenth of $\rm kV/cm$, the photon frequencies are in the terahertz
range.

In fact, semi-classical calculations exhibit neither gain nor
absorption at resonance, \emph{i.e.} for $\omega = \omega_b$, but
transparency. Particularly in the quantum-mechanical picture, it
is evident that only spontaneous transitions can occur at
resonance. At sufficiently high electric fields, the miniband is
split into the Wannier-Stark ladder, a set of states evenly spaced
by $eFd$ in energy. Resonant stimulated emission processes between
adjacent states are balanced by absorption processes, because of
the translational symmetry of the system, which dictates equal
occupation for all rungs of the Wannier-Stark ladder.

Nevertheless, the semi-classical calculation does predict gain --
without inversion -- for non-resonant transitions with a "too
small" photon energy, $\hbar\omega<\hbar\omega_b$, and absorption
for $\hbar\omega>\hbar\omega_b$. But the existence and strength of
this semi-classically predicted Bloch gain is still discussed,
despite experimental observation of many related phenomena in
superlattices, such as negative differential conductivity
\cite{Sibille:PRL:90:52}, the associated Bloch oscillations
\cite{Feldmann:PRB:92:7252} and the coupling of the superlattice
to external THz radiation
\cite{Keay:PRL:95:4102,Unterrainer:PRL:96:2973}, just to mention a
few. In particular, the gain mechanism is lacking an
interpretation in terms of the quantum-mechanical Wannier-Stark
picture.

In this paper such a quantum-mechanical interpretation of the
Bloch gain in superlattices is suggested and a link is established
between the intersubband gain originating from a population
inversion, with its symmetric spectral shape centered at the
transition energy, and the dispersive gain predicted for a
periodic superlattice, with its nearly anti-symmetric profile. The
quantum-mechanical model, based on the density matrix formalism
similar to the one employed earlier \cite{Kazarinov:SPS:72:120},
yields a general expression for the gain profile in intersubband
transitions.

In Sec.~\ref{sec2} we present the model system, and discuss
assumptions and details of the density matrix calculation. With
less stringent approximations than those made by Kazarinov and
Suris we find an expression for the coherence between two, at
first spatially separated, subband states that are coupled by
tunneling and broadened by intra-subband scattering. The coherence
determines current density as well as optical transitions.
Transforming to the basis of eigenstates of the biased
heterostructure -- the Wannier-Stark basis for the superlattice --
the model is extended to describe optical transitions between any
pair of subbands.

In Sec.~III we apply the theory to superlattices and obtain the
quantum mechanical counterparts to the semi-classical results for
both the Esaki-Tsu current-voltage characteristics and the
dispersive Bloch gain. The results are quantitatively compared to
the predictions of the semi-classical picture, where good
agreement is found for higher electron temperatures.

In Sec.~IV intersubband transitions are investigated for the
quantum cascade structure and the relation between
(anti-symmetric) Bloch gain and Lorentzian-shaped intersubband
gain becomes apparent: the theory predicts a transition from the
Lorentzian shaped inversion gain to the dispersive Bloch gain with
decreasing population inversion accompanied by a red-shift of the
peak gain with respect to the transition energy.

\section{Theory}\label{sec2}

\begin{figure}[htb]
\includegraphics[width=3.3in]{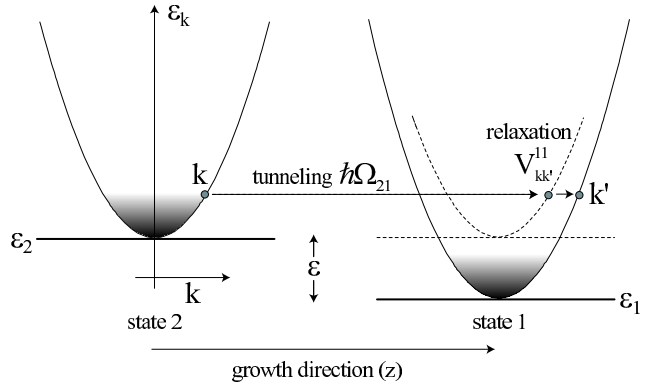} \caption{Mixed momentum and real-space
picture of a two-level-system that serves as a simple model for a
diagonal intersubband transition. Tunneling into a virtual
intermediate state (dotted) as well as photon-assisted tunneling
is expressed by a transfer matrix element $\hbar\Omega_{ij}$.
Relaxation is assumed to take place within each subband
only.}\label{schema_intro}
\end{figure}

We consider two subbands, confined in adjacent wells, that serve
as a model system for photon-assisted tunneling structures and, in
particular, for transitions within the Wannier-Stark ladder of a
weakly coupled superlattice. To start with, the same basis as in
the original work of Kazarinov and Suris is chosen as an
appropriate basis set. The wave functions $\Ket{ik}$ are given by
the product of the envelope functions $\Psi_i(z)$, maximally
localized \cite{Kohn:PRL:59:809} in well $i$, and plane waves
\begin{equation}
\BraKet{{\bf r},z}{ik} = \Psi_i(z)\,e^{i{\bf k\cdot r}}\,,
\end{equation}
where the $z$-axis is defined by the direction of growth, ${\bf k}
= (k_x,k_y)\equiv k$ denotes the in-plane momentum and ${\bf
r}=(x,y)$ the lateral position. The matrix elements of the
Hamiltonian in this basis are given by
\begin{equation}
\mathcal{H}_{kk'}^{ij} = \BraOpKet{ik}{\mathcal{H}}{jk'} =
H_{k'}^{ij}\delta_{kk'} + V_{kk'}^{ij}\,,
\end{equation}
where the respective contributions $H$ and $V$ take the form
\begin{equation}
H^{ij}_k= \left(\begin{array}{ccc} \epsilon_{2k} &
\hbar\Omega_{21}\\ \hbar\Omega_{12} &
\epsilon_{1k}\\\end{array}\right)^{ij},\;V^{ij}_{kk'}=
\left(\begin{array}{ccc} V_{kk'}^{22} & 0 \\ 0 & V_{kk'}^{11}\\
\end{array} \right)^{ij} \nonumber \,.\end{equation}
Thus, electrons are allowed to tunnel between the subband state
$i$ and $j$ by means of the momentum conserving matrix elements
$\hbar\Omega_{ij}$, in each of which they are possibly scattered
out of a virtual intermediate state by an intra-well relaxation
process $V_{kk'}^{ii}$ as depicted in Fig.~\ref{schema_intro}. For
simplicity we restrict ourselves to elastic (impurity,
\emph{e.g.}) scattering within each subband and assume a parabolic
dispersion relation parallel to the layers in the effective mass
approximation
\begin{equation}
\epsilon_{ik} = \epsilon_i+\frac{\hbar^2 k^2}{2m^*}\,,
\end{equation}
where $\epsilon_i$ denotes the lower subband edge and $m^*$ is the
effective mass of the electron averaged over the extension of the
wave function in well and barrier.

\subsection{Coherences}

Using the equation of motion of the density matrix
$i\hbar\,\partial_t\hat{\rho}=[\mathcal{H},\hat{\rho}]\label{timeevo}$
and separating the diagonal and non-diagonal part with respect to
the parallel momentum $k,k'$ according to
$\hat{\rho}^{ij}_{kk'}=\delta_{kk'}\rho^{ij}_{k}+(1-\delta_{kk'})\rho^{ij}_{kk'}\label{splitting}\,,$
we obtain two coupled equations with four terms each, which
determine the time evolution of the system. Since the coherent
term $H$ is diagonal with respect to the in-plane momentum and the
scattering term $V$ is purely non-diagonal, the commutators that
determine the diagonal and non-diagonal part of the density matrix
are evaluated as
\begin{eqnarray}
i\hbar\,\partial_t\rho^{ij}_{k}&& =
\sum_{m}(H_{k}^{im}\rho^{mj}_{k}-\rho^{im}_{k}H_{k}^{mj})\nonumber\\&&
+\sum_{m,k'}(V_{kk'}^{im}\rho^{mj}_{k'k}-\rho^{im}_{kk'}V_{k'k}^{mj})\\
\label{25} i\hbar\,\partial_t\rho^{ij}_{kk'}&&\approx
\sum_{m}(V_{kk'}^{im}\rho^{mj}_{k'}-\rho^{im}_{k}V_{kk'}^{mj})\nonumber\\&&+
\sum_{m}(H_{k}^{im}\rho^{mj}_{kk'}-\rho^{im}_{kk'}H_{k'}^{mj})\,,\label{26}
\end{eqnarray}
where the commutator of the scattering potential with the
non-diagonal part of $\rho$ has been neglected in the second
equation (Born approximation). The steady-state values of the
coherences of the density matrix $f^{ij}$, which determine the
transitions $\Ket{ik}\rightarrow\Ket{jk}$, the current and the
absorption, are obtained from a Laplace average
\cite{Kohn:PRL:57:590} defined by
\begin{equation}
f(s)=s\int_0^\infty dt e^{-st}\rho(t)
\end{equation}
and performing the Laplace limit $s\rightarrow0$ using the
relation
$\lim_{s\rightarrow0_+}(\omega-is)^{-1}=P(1/\omega)+i\pi\delta(\omega)$
at the appropriate stage of the calculation. In this approach the
populations $f^{ii}_k$ of the density matrix are not accessible
and appear in the resulting expression as external quantities. The
Laplace average gives
\begin{eqnarray}
i\hbar s f^{ij}_{k}&& = i\hbar s\rho^{ij}_{k}(0) +
\sum_{m}(H_{k}^{im}
f^{mj}_{k}-f^{im}_{k}H_{k}^{mj})\nonumber\\&&+\sum_{m,k'}(V_{kk'}^{im}f^{mj}_{k'k}-f^{im}_{kk'}V_{k'k}^{mj})\label{25f}\\i\hbar
s f^{ij}_{kk'}&&\approx i\hbar s \rho^{ij}_{kk'}(0)+
\sum_{m}(V_{kk'}^{im} f^{mj}_{k'}-
f^{im}_{k}V_{kk'}^{mj})\nonumber\\&&+ \sum_{m}(H_{k}^{im}
f^{mj}_{kk'}- f^{im}_{kk'}H_{k'}^{mj})\label{26f}\,.
\end{eqnarray}
In a first step, the non-diagonal part in equation (\ref{26f}) is
approximated (\emph{cf.} appendix A for details). Specifying to
the assumptions of a two-level-system with intra-well scattering
only and neglecting terms of higher order in $\Omega_{ij}$
corresponding to multiple tunneling processes, gives
\begin{eqnarray}
\left(f\right)^{ij}_{kk'}&&\approx i\pi\delta(\epsilon_{ik}
-\epsilon_{jk'})(V_{kk'}^{ii} f^{ij}_{k'}-f^{ij}_{k}V_{kk'}^{jj}+
\label{approx}\\&&+\hbar\Omega_{ij}\left(\frac{V_{kk'}^{jj}(f^{jj}_{k}-
f^{jj}_{k'})}{\epsilon_{jk} -\epsilon_{jk'}}-
\frac{V_{kk'}^{ii}(f^{ii}_{k}- f^{ii}_{k'})}{\epsilon_{ik}
-\epsilon_{ik'}}\right))\nonumber\,,\end{eqnarray} which has to be
placed into equation (\ref{25f}) for the diagonal part.
Simplifying for intra-well scattering here and taking the Laplace
limit yields (\emph{cf.} appendix B)
\begin{eqnarray}
(\epsilon_{ik} -\epsilon_{jk})&&f^{ij}_{k} =
\hbar\Omega_{ij}(f_{k}^{ii} - f_{k}^{jj}) \nonumber\\&&-
\sum_{k'}(V_{kk'}^{ii}\left(f\right)_{k'k}^{ij}
-\left(f\right)_{kk'}^{ij} V_{k'k}^{jj})\,,
\end{eqnarray}
where ($f$) denotes the approximated expression for the
non-diagonal part in equation (\ref{approx}). Performing an
ensemble average, \emph{i.e.} dropping terms related to
correlation effects in the scattering potential, we obtain
\begin{eqnarray}
&& (\epsilon_{ik}-\epsilon_{jk}) f^{ij}_{k} - i\pi
f^{ij}_{k}\times\nonumber\\
&&\times\sum_{k'}\delta(\epsilon_{ik'}-\epsilon_{jk})|V_{kk'}^{ii}|^2
+\delta(\epsilon_{ik}-\epsilon_{jk'})|V_{kk'}^{jj}|^2\nonumber\\
&&= \hbar\Omega_{ij}(f_{k}^{ii} - f_{k}^{jj})+
\nonumber\\&&+i\pi\sum_{k'}\bigl[
\delta(\epsilon_{ik'}-\epsilon_{jk})|V_{kk'}^{ii}|^2
\frac{\hbar\Omega_{ij}}{\epsilon_{ik'}-\epsilon_{ik}}
(f^{ii}_{k'}-f^{ii}_{k}) \nonumber\\
&&+\delta(\epsilon_{ik}-\epsilon_{jk'})|V_{kk'}^{jj}|^2
\frac{\hbar\Omega_{ij}}{\epsilon_{jk}-\epsilon_{jk'}}
(f^{jj}_{k}-f^{jj}_{k'})\bigr]\,,
\end{eqnarray}
which agrees with the previous result \cite{Kazarinov:SPS:72:120}.
In contrast to the original treatment, we continue by neither
neglecting the difference of the arguments in the
$\delta$-functions on the LHS nor omitting the second term on the
RHS. The coherence associated with the transition
$\Ket{2k}\rightarrow\Ket{1k}$ is obtained from
\begin{eqnarray}
&&\epsilon f_k^{21}
-\overbrace{i(\gamma_{k}^{2}+\gamma_{k}^{1})}^{\text{transition
broadening}}f_k^{21}=
\overbrace{\hbar\Omega_{21}(f_k^{22}-f_k^{11})}^{\text{population
difference}}\nonumber\\
&&+\underbrace{i\hbar\Omega_{21}\epsilon^{-1}(\gamma_{k}^{2}(f_{q_-}^{22}-f_k^{22})
-\gamma_{k}^{1}(f_{q_+}^{11}-f_k^{11}))}_{\text{Bloch type
contribution}}\,,
\end{eqnarray}
where we have used abbreviations for the scattering induced
broadening of the transition
$\gamma_k^{i}=\sum_{k'}\delta(\epsilon_{ik'}-\epsilon_{jk})|V_{kk'}^{ii}|^2$,
the subband separation $\epsilon=\epsilon_{2k}-\epsilon_{1k}$ and
the in-plane momentum of the final state $q_\pm =
\hbar^{-1}\sqrt{2m^*(\epsilon_k\pm\epsilon)}$ . The first term on
the RHS, which contains the difference of populations between the
two states, corresponds to the central result of Reference
\cite{Kazarinov:SPS:72:120}, cited often as the original proposal
of the quantum cascade laser. The second term, which has been
discarded so far, contains differences in population within a
subband. It is this term which will be responsible for the second
order type of gain, leading to the characteristic negative
differential conductivity and the dispersive gain profile in a
superlattice, and a modified spectral shape of the gain in a
quantum cascade laser.

\subsection{Current density}

The current density between two states spatially separated by
$d=z_{22}-z_{11}$ is calculated from $j = e\text{Tr}(\hat{v}f)$,
where $\hat{v}=i/\hbar[H,z]$ is the velocity operator and $z$ is
the position operator. The current is induced by the non-diagonal
matrix elements of the velocity operator $v_{ij}$, which are given
by $v_{ij} = i\Omega_{ij}(z_{jj}-z_{ii})+\epsilon z_{ij}/\hbar$.
By choice of the basis set, the contribution of the dipole
$z_{ij}$ is small compared to the tunneling term
\cite{Kazarinov:SPS:72:120} and we obtain
\begin{eqnarray}
&&j \approx e d \sum_k i(\Omega_{21}
f_{k}^{12}-\Omega_{12}f_{k}^{21})\,.
\end{eqnarray}
Using the previous equation for the coherences $f_k^{12}$ and
$f_k^{21}$ and the current yields
\begin{eqnarray}
j=\frac{ed|\hbar\Omega_{21}|^2}{\hbar}\sum_k\frac{\gamma^1_k
(f^{22}_{k}-f^{11}_{q_+})+\gamma^2_k
(f^{22}_{q_-}-f^{11}_{k})}{\epsilon^2
+(\gamma_k^1+\gamma_k^2)^2}\,.\label{current}
\end{eqnarray}
The current results from differences in population. In the
following section, and in contrast to the original work, the
differences are evaluated for non-equivalent $k$-states in the
respective subbands. To obtain the result of Kazarinov and Suris,
$q_\pm$ is set equal to $k$ and a constant broadening $\gamma$ is
used
\begin{eqnarray}
j\approx\frac{ed|\hbar\Omega_{21}|^2}{\hbar}\frac{\gamma}{\epsilon^2
+\gamma^2}\Delta n\,.\nonumber
\end{eqnarray}
With this approximation, the current density is solely driven by
the density of excess electrons in either state $\Delta
n=\sum_k(f^{22}_{k}-f^{11}_{k})$. For a superlattice, this
approximation does predict the resonant current peaks that occur
whenever ground and excited states align, but fails to account for
the current between equivalent states in the Wannier-Stark ladder.

\subsection{Absorption and gain}

Optical properties are deduced from the high-frequency response to
an additionally applied ac field. In the case of a photon-assisted
tunneling transition, the Hamiltonian is supplemented by
\begin{equation}
\delta H(t)= -\frac{e}{c}\,\hat{v}A = \left(\begin{array}{ccc} 0 &
\frac{edf_\omega}{\omega}\Omega_{21}e^{-i\omega t}\\
\frac{edf_\omega}{\omega}\Omega_{12}e^{i\omega t} &
0\\\end{array}\right)\end{equation} for a vector potential $A =
(c/i\omega) f_\omega e^{-i\omega t}$ with an amplitude $f_\omega$
of the high-frequency field. Noting the similar structure of
$\delta H$ and the non-diagonal part of $H$, the corrections to
the coherences in linear response $\delta f_k^{21}$ and $\delta
f_k^{12}$ are related to $f_k^{21}$ and $f_k^{12}$ by
\begin{equation} \delta f_{k,\omega}^{21} = -
\frac{edf_\omega}{\hbar\omega}f^{21}_{k,\epsilon-\hbar\omega}
\;\;\text{and}\;\; \delta f_{k,\omega}^{12} =
\frac{edf_\omega}{\hbar\omega}f^{12}_{k,\epsilon+\hbar\omega}\,,
\end{equation}
which are evaluated at an energy $\epsilon\pm\hbar\omega$ instead
of $\epsilon$ due to the time dependence of the ac field. This
relation reflects the similarity of tunneling and photon-assisted
tunneling processes in a diagonal transition. The photon-induced
current becomes
\begin{eqnarray}
\delta j(\omega) \approx e d \sum_k i(\Omega_{21} \delta
f_{k,\omega}^{12}-\Omega_{12} \delta f_{k,\omega}^{21})\,.
\end{eqnarray}
The high-frequency conductivity is related to the current by
$\sigma(\omega)=\partial(j+\delta j(\omega))/\partial f_\omega$,
and directly linked to the absorption (\emph{cf.} appendix C for
details) according to
\begin{eqnarray}\label{diagonal}
\alpha(\omega) &&= \frac{\Re(\sigma(\omega))}{\varepsilon_0n_rc} =
- \frac{e^2 d^2 |\Omega_{21}|^2}
{\varepsilon_0n_rc\omega}\label{diag} \times\\ \times&&\sum_k
\frac{\gamma_{k}^1(f_k^{22}-f^{11}_{k_+})
+\gamma_{k}^2(f^{22}_{k_-}-f_k^{11})}
{(\epsilon-\hbar\omega)^2+(\hbar\tau_k^{-1})^2})\nonumber\,.
\end{eqnarray}
The in-plane momenta of the final states are denoted by $k_\pm =
\hbar^{-1}\sqrt{2m^*(\epsilon_k\pm(\epsilon-\hbar\omega))}$. The
preceding expression contains the two gain mechanisms as limiting
cases of a more general intersubband gain profile with a simple
physical interpretation. Before we discuss the quantum-mechanical
paths involved, the expression may be generalized to an
arbitrarily located pair of subband states.

\begin{figure}[htb]
\includegraphics[width=2.2in]{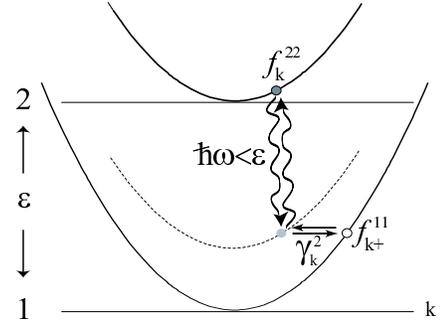}
\caption{Possible quantum-mechanical paths: for an incident photon
with energy $\hbar\omega\neq\epsilon$, absorption or stimulated
emission may occur due to a non-resonant absorption or emission
into an intermediate and a subsequent relaxation into the final
state. Energy and momentum are conserved in this second order
process.}\label{schema_paths}
\end{figure}

To account for vertical as well as diagonal transitions the basis
of eigenstates of the biased system -- the Wannier-Stark basis for
a superlattice -- is chosen. Due to the assumption of intra-well
scattering only, the dark current vanishes as the tunneling matrix
element $\hbar\Omega_{ij}$ is incorporated in the extended wave
functions. The photon-induced current, however, is then mediated
by the dipole matrix element between the two subband states, which
can no longer be considered as small in this basis. The
non-diagonal part of the velocity operator is given by $v_{ij} =
i\epsilon z_{ij}$. Since the operator of the high frequency field
$\delta H = -e/c \hat{v}A$ does not change its purely non-diagonal
structure, inspection of the previous equations and replacement of
$id\Omega_{ij}$ by $i\epsilon z_{ij}/\hbar$ naturally extends the
equation for the gain profile to any intersubband transition and
permits to omit the rather arbitrary distinction between a
diagonal and vertical transition
\begin{equation}\alpha(\omega) = -\frac{e^2
|z_{21}|^2\epsilon^2 }{\varepsilon_0n_rc\hbar^2\omega} \sum_k
\frac{\gamma_{k}^1(f_k^{22}-f^{11}_{k_+})+\gamma_{k}^2(f^{22}_{k_-}-f_k^{11})}
{(\epsilon-\hbar\omega)^2+(\gamma_k^{1}+\gamma_k^{2})^2}\,.\label{gain}
\end{equation}
As will be shown in the following, expression (\ref{gain}) allows
a simple explanation of the gain mechanism in a superlattice and
in a quantum cascade structure. It is instructive to rewrite the
differences in populations as
\begin{eqnarray}
\gamma_{k}^1(f^{22}_{k}-f_{k_+}^{11}) =
\gamma_{k}^1\underbrace{(f^{22}_{k}(1-f_{k_+}^{11})}_{
\text{emission}\Ket{2}\rightarrow \Ket{1}}
-\underbrace{f_{k_+}^{11}(1-f^{22}_{k}))}_{
\text{absorption}\Ket{1}\rightarrow \Ket{2}},
\end{eqnarray}
which directly translate into the paths depicted in
Fig.~\ref{schema_paths}. The two processes above relate the states
$\Ket{2k}$ and $\Ket{1k_+}$ by the emission or absorption of a
(non-resonant) photon, $\hbar\omega \neq \epsilon$ for $k \neq
k_+$, assisted by relaxation within the lower state via
$\gamma_k^1$, which ensures momentum transfer. The second
difference in equation (\ref{gain}) is interpreted accordingly,
where the relaxation takes place within the upper state.

If one assumes constant and equal in-plane scattering times
$\tau=\hbar/\gamma$, a Fermi distribution with the same
temperature $T$ in each subband and chemical potentials $\mu_1$
and $\mu_2$, respectively, the gain is analytically expressed as
\begin{eqnarray}
\alpha(\omega)&&=\frac{e^2 |z_{12}|^2\epsilon^2
}{\varepsilon_0n_rc\hbar^2\omega} \frac{m^* k_b
T}{\pi\hbar^2}\frac{\gamma}{\delta^2+\gamma^2}\times\nonumber\\&&
\times\ln\left(\frac{e^{(\mu_1-\epsilon_1)/k_b T}e^{-\delta/k_b
T}+\hat{\theta}(\delta)}{e^{(\mu_2-\epsilon_2)/k_b
T}+\hat{\theta}(\delta)}\right)\,,
\end{eqnarray}
where $\delta = \epsilon-\hbar\omega$ characterizes the
off-resonant nature of the photon transition and
$\hat{\theta}(\delta)=\theta(\delta)+\theta(-\delta)e^{-\delta/k_b
T}$ reflects the asymmetry between "too small" and "too large"
photons with regard to the resonant transition.

\section{Results}

In this part the theoretical model is evaluated and interpreted
with respect to the gain profile of a superlattice and quantum
cascade laser.

\subsection{Bloch oscillator}

In the case of a superlattice, populations and scattering times
are equal for symmetry reasons, $f_k^{11}=f_k^{22}$ and
$\gamma_k^{1}=\gamma_k^{2}$. The characteristic negative
differential conductivity of the current-voltage characteristic
$j(F)$ is recovered from equation (\ref{current}) if one
identifies the subband spacing with the field drop per period,
$\epsilon=eFd$, where $F$ is the applied electric field. The
current density reads
\begin{eqnarray}
j(F) =\frac{ed\Delta^2}{4\hbar}\sum_k\frac{\gamma_k
(f_{k_-}-f_{k_+})}{(eFd)^2 +(2\gamma_k)^2} \,,\label{ivesaki}
\end{eqnarray}
where we omitted the state indices and introduced the miniband
width via $\Delta\approx 4\hbar\Omega_{12}$. The populations may
be described by thermal distributions, either by Fermi-Dirac or
Boltzmann statistics. The current-voltage characteristic resembles
the Esaki-Tsu characteristic and agrees quantitatively with the
result by Wacker \emph{et al.} \cite{Wacker:ttsn:98:321} within
the sequential tunneling picture \cite{Wacker:PRB:97:13268} valid
for weakly coupled superlattices. This assumption is implicit in
the present approach, as we allow only for next-neighbor
interaction and multiple tunneling is excluded, corresponding to a
limited coherence of spatially extended states.

\begin{figure}[htb]
\includegraphics[width=3.3in]{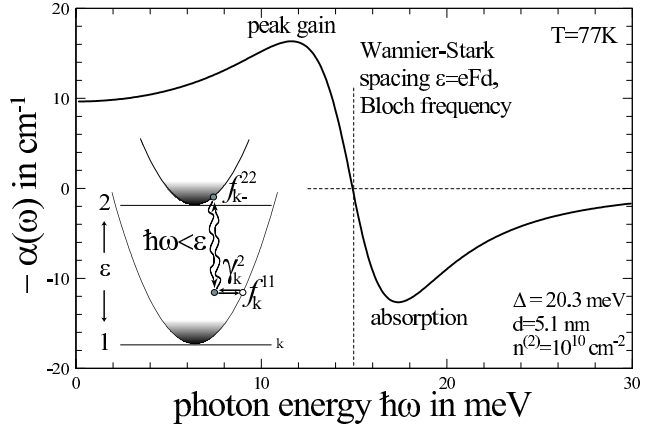}
\caption{Bloch oscillator at $\epsilon=eFd=15\rm meV$: a
dispersive gain contribution arises at $\Delta n = 0$ from second
order processes by non-resonant photon emission/absorption
followed by scattering events that ensure conservation of
momentum. Whereas stimulated emission is predicted for
$\hbar\omega<\epsilon$ (\emph{cf.} Inset: path of stimulated
emission), absorption dominates for $\hbar\omega>\epsilon$. A
constant in-plane relaxation time of $\tau=0.2\rm\,ps$ is used.
Parameters of the GaAs/AlAs-superlattice \cite{Rott:PRB:99:7334}:
$\Delta=20.3\rm meV$, $d=5.1\rm nm$,
$n^{\left(2\right)}=10^{10}\rm cm^{-2}$. }\label{schema_bo}
\end{figure}

Rewriting the gain profile of equation (\ref{gain}) specifically
for a superlattice yields
\begin{equation}
\alpha(\omega)=
-\frac{e^2d^2|\frac{1}{4}\Delta|^2}{\varepsilon_0n_rc\hbar^2\omega}
\sum_k \frac{\gamma_{k}(f_{k_-}-f_{k_+})}
{(eFd-\hbar\omega)^2+(2\gamma_k)^2}\,,\label{superlattice}
\end{equation}
Note, as the miniband width of a superlattice, $\Delta$, and the
dipole matrix element, $z_{ij}$, are related by
\cite{Unterrainer:ssv66:00:4} $z_{ij} = d \Delta/4eFd$ in the
Wannier-Stark basis, the diagonal expression of equation
(\ref{diagonal}) and the general expression of equation
(\ref{gain}) for $\alpha(\omega)$ provide identical results. At
resonance, the incoming photon provokes transitions between
equivalent states, $k_\pm = k$, and absorption and emission
balance each other as expected from a system with no population
inversion, $\Delta n= \sum_k (f^{22}_k-f^{11}_k)=0$. In the case
of photons with energy $\hbar\omega < \epsilon$ the lower state
involved in this second order transitions will be less occupied
than the upper state, leading to an asymmetry between emission and
absorption in favor of gain. In contrast, for a photon energy
exceeding the subband spacing absorption occurs. As illustrated in
Fig.~\ref{schema_bo} equation (\ref{superlattice}) recovers the
dispersive shape of the Bloch gain -- giving rise to absorption
above and (stimulated) emission below the field-dependent Bloch
frequency $\omega_b=eFd/\hbar$. Note, we have neglected any
particularity of the actual scattering processes here. A detailed
investigation in the framework of second order perturbation theory
\cite{Schmidt:unpublished} reveals a complex interplay of
population effects and the influence of the momentum transfer for
the relaxation processes in a superlattice.

We compare the Bloch gain derived within the present approach with
the results obtained from the standard model based on
semi-classical calculations. In the semi-classical approach
\cite{Ktitorov:FTT:71:2230} the Boltzmann equation is solved in
the relaxation time approximation for the distribution function
$f(k_z,k_\parallel)$ of miniband electrons subject to an external
dc and ac field $F(t)=F+F_\omega \cos(\omega t)$, where
$F_\omega\ll F$. In the case of a Maxwell distribution
\cite{Ignatov:PSSB:76:327} this yields
\begin{eqnarray}
\alpha_{sc}(\omega)&&= \frac{e^2 d^2}{\varepsilon_0 n_r
c}\,\frac{\Delta}{2\hbar^2}\;n^{(3)}\frac{I_1(\Delta/2k_BT)}
{I_0(\Delta/2k_BT)}\times\\ \times&&
\frac{\tau}{1+(\omega_b\tau)^2}
\Re\left(\frac{1-i\omega\tau-(\omega_b\tau)^2}{(\omega_b\tau)^2+
(1-i\omega\tau)^2}\right)\nonumber\,,\label{ktitorov}
\end{eqnarray}
in the single relaxation time approximation and
\begin{eqnarray}\label{ignatov}
&&\alpha_{sc}(\omega) = \frac{e^2 d^2}{\varepsilon_0 n_r
c}\,\frac{\Delta}{2\hbar^2}\;n^{(3)}\frac{I_1(\Delta/2k_BT)}
{I_0(\Delta/2k_BT)}\times\\ \times&&
\frac{\tau_p}{1+\omega_b^2\tau_e\tau_p}
\Re\left(\frac{1-i\omega\tau_e-\omega_b^2\tau_e\tau_p}
{\omega_b^2\tau_e\tau_p+(1-i\omega\tau_e)(1-i\omega\tau_p)}
\right)\nonumber\,,
\end{eqnarray}
for the improved two relaxation time approximation given by
Ignatov \emph{et al.}, where distinct momentum and energy
relaxation times $\tau_p$, $\tau_e$ are used and which agrees with
detailed Monte-Carlo studies \cite{Willenberg:montecarlo}. The
ratio of Bessel functions contains the temperature dependence for
a non-degenerate electron gas.
\begin{figure}[htb]
\includegraphics[width=3.3in]{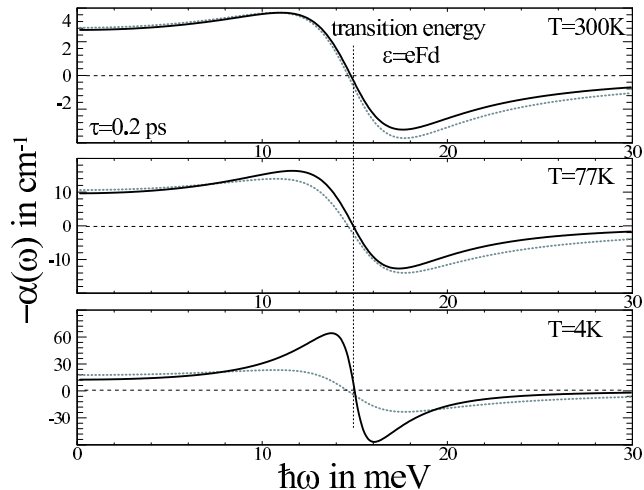}
\caption{Semi-classical (dotted line) vs quantum-mechanical
results (full line) for the absorption in a superlattice for
different temperatures $T$. We assume a temperature-independent
scattering time $\tau=0.2\rm\,ps$ in the quantum-mechanical model
and set $\tau_{k,e}=\tau$ in the semi-classical model. In the
semi-classical picture the peak gain scales with the ratio
$I_1(\Delta/2k_BT)/I_0(\Delta/2k_BT)$. The quantum mechanical gain
profile exhibits an additional narrowing with lower
temperature.}\label{comparison}
\end{figure}
Fig.~\ref{comparison} shows a comparison of the semi-classical
results and the quantum mechanical predictions for the same
constant relaxation time, $\tau=\hbar/\gamma=0.2\rm\,ps$, at
different temperatures $T$. No independent parameters are used.
The two approaches agree remarkably well at high temperatures in
the semi-classical limit $eFd<\Delta$. The narrowing of the Bloch
gain profile with lower temperature, compared to the
semi-classical curve, reflects an explicit influence of the
electron distribution within the subband. This influence is absent
in the semi-classical treatment, regardless of the approximation
for the distribution function. In real devices, however, the
electron temperature reaches $100\rm\,K$ and above, if the
superlattice is biased beyond the onset of Bloch oscillations,
according to a self-consistent theoretical analysis of the
in-plane distribution function in the Wannier-Stark picture
\cite{Rott:PRB:99:7334}. Still, at electron temperatures of about
$77\rm\,K$, the considered superlattice, \emph{e.g.}, with a sheet
density of $n^{(2)}=10^{10}\rm\,cm^{-2}$ exhibits a peak material
gain of about $15\,\rm cm^{-1}$, which exceeds the estimated value
for waveguide losses in the terahertz range
\cite{Rochat:APL:02:submitted}.

\subsection{Quantum cascade laser}

\begin{figure}[htb]
\includegraphics[width=3.3in]{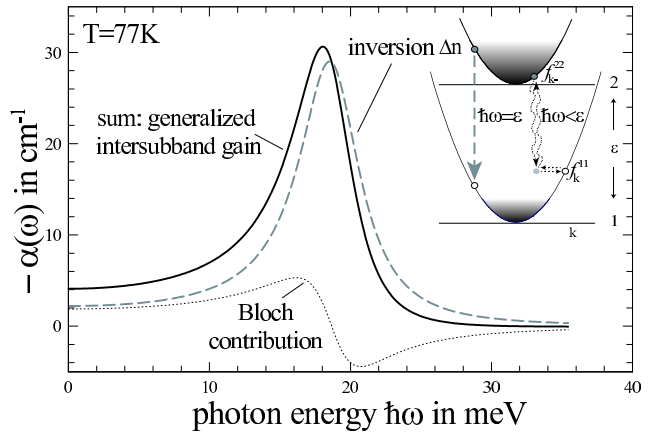}
\caption{The population inversion, $\Delta n$, leads to nearly
symmetric gain profile centered at $\hbar\omega = \epsilon$
(\emph{cf.} left hand side of Inset, dashed line). In addition,
the dispersive Bloch gain contributes also here (\emph{cf.} right
hand side of Inset, dotted line). The generalized intersubband
gain consists of both contributions. Its spectral shape becomes
more and more asymmetric with decreasing population inversion. The
sample parameters correspond to the terahertz quantum cascade
lasers \cite{Koehler:Nat:02:156, Rochat:APL:02:submitted} at about
$\hbar\omega=18.7\rm\,meV$. A constant relaxation time
$\tau=0.5\rm\,ps$ and populations $n_2=3\cdot 10^{9}\rm\,cm^{-2}$,
$n_1=1\cdot10^{9}\rm\,cm^{-2}$ are used, $\Delta n/n \approx
0.5$.}\label{schema_qcl}
\end{figure}

\begin{figure}[htb]
\includegraphics[width=3.3in]{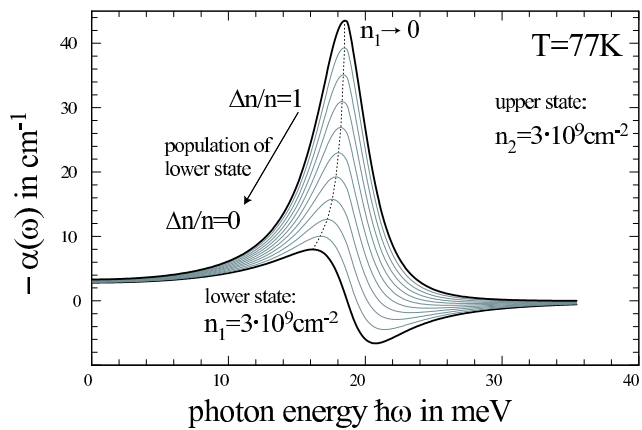}
\caption{Evolution of the generalized gain profile from the
inversion gain - linearly depending on the population inversion -
to the red-shifted Bloch gain with decreasing $\Delta n/n$,
respectively increasing the lower state population $n_1$, while
keeping the upper state density $n_2$ constant.\label{evolution}}
\end{figure}

In the quantum cascade laser, the populations of the respective
subband state depend on the design as well as current and
temperature. Then, the inversion gain and a Bloch type
contribution add up as shown in Fig.~\ref{schema_qcl}. For a
negligible lower state population, $\Delta n/n\approx 1$, where
$n=\sum_k (f^{22}_k+f^{11}_k)$, equation (\ref{gain}) is dominated
by resonant photon emission due to the population inversion
between equivalent $k$-states. It resembles the Lorentzian shaped
inversion gain profile, linearly depending on the population
inversion $\Delta n$.
On the other hand, in the limiting case of equal populations, the
Bloch type contribution results in a dispersive gain profile as in
the superlattice. In between, there is a smooth transition of the
(usual) intersubband gain profile to the dispersive Bloch gain
with decreasing $\Delta n/n$ as shown in Fig.~\ref{evolution}.

Thus, equation (\ref{gain}) for the gain profile states, that
\emph{there is a dispersive contribution to the gain profile in
any intersubband transition}, with a rising significance for
$\Delta n/n$ tending to zero. This result implies two predictions
for an intersubband emitter such as the quantum cascade laser.
First, the gain does not linearly depend on $\Delta n$, but there
is a non-negligible intersubband gain even without a population
inversion, scaling approximately linearly with the electron
density $n$ in the system. Secondly, above threshold, the peak
gain and, thus, the laser signal is expected to shift to lower
energies with increasing current or temperature, as the ratio
$\Delta n/n$ generally decreases.

\section{Discussion}

In sections II and III of this paper, the homogeneous broadening
$\gamma^i_k$ of intrasubband relaxation processes has been
introduced, without specifying it in detail. For the numerical
evaluations it was taken as $k$-independent quantity. As mentioned
before, a realistic calculation for microscopic interaction
processes in a superlattice will be given elsewhere
\cite{Schmidt:unpublished}. Moreover, the effect of inhomegeneous
level broadening due to interface roughness has not yet been
discussed. Inhomogeneity can be considered in a simplified
approach such as the "local quantum-mechanical model"
\cite{Metzner:PRB:95:5106,Willenberg:PRB:02:35328}. This model
assumes a "global quasi-Fermi-level" in each subband, independent
of the in-plane position. Within this model the dispersive shape
of the Bloch gain is not obscured by inhomogeneous broadening.
This holds true even in a diagonal structure
\cite{Willenberg:disorder}, where the subband fluctuations are not
correlated, though the line width will be determined by
inhomogeneous broadening \cite{Campman:APL:96:2554}. Furthermore,
intersubband plasmons are known to alter the line shape of the
intersubband transition and to cause a blue-shift of the
intersubband resonance due to dynamical screening of the dipole
field \cite{Warburton:SM:96:365}. However, the subband states of a
Bloch oscillator or quantum cascade laser are generally weakly
populated $\approx 10^{10}\rm cm^{-2}$ compared to the onset of
the collective phenomena \cite{Luin:PRB:01:041306-1} beyond some
$10^{11}\rm cm^{-2}$.

In conclusion, the proposed model provides a unified description
of optical transitions between two-dimensional subbands. The upper
and lower subband can be either of the same kind (superlattice) or
of different kind (quantum cascade laser structure).

In a superlattice a population inversion between equivalent states
differing in energy by  $\epsilon=eFd$ cannot occur, as
$f^{22}_k=f^{11}_k$, due to the translational symmetry of the
system. Consequently, resonant stimulated photon emission
processes are exactly balanced by the corresponding absorption
processes. However, non-resonant second order processes exhibit
gain for $\hbar\omega<\epsilon$, whereas net absorption occurs for
$\hbar\omega>\epsilon$. This inversionless gain at
$\hbar\omega<\epsilon$ represents the quantum mechanical analogue
to the Bloch gain predicted by semi-classical models, which had
not been described previously. The quantum-mechanical approach
agrees remarkably well with the semi-classical results in the
high-temperature limit. In contrast to the miniband picture
\cite{Kroemer:CondMat:00:xx}, it provides an easily conceivable
interpretation of the gain mechanism.

In a quantum cascade laser structure, upper and lower state
generally exhibit a different population, and in the ideal case,
an inverted population. If the inversion decreases, the
quasi-symmetric gain spectrum at a high degree of inversion, where
$\Delta n/n\approx 1$, evolves to the dispersive Bloch gain, for
$\Delta n/n\approx 0$ and below. In contrast to the peak inversion
gain, which does not depend on temperature and decreases with
scattering, the latter decreases with temperature as the
differences in occupation between initial and final states
diminish, but increases with more frequent scattering processes.

The theory predicts amplification without inversion below the
intersubband resonance of two broadened states. The peak gain in
any amplified intersubband transition relying on a poor population
inversion, \emph{i.e.} $\Delta n/n\approx 0$, exhibits a red-shift
of the order of the level broadening $\gamma$ with respect to the
transition energy. This dispersive gain contribution, that is
responsible for the Bloch oscillator effect and which escaped
observation so far, is expected to be experimentally accessible in
a quantum cascade structure by a search for the attributed
red-shift.

\section*{Acknowledgements}

We would like to thank Andreas Wacker, Daniel K\"orner and Giacomo
Scalari for fruitful discussions. This work was supported by the
Swiss National Science Foundation.



\begin{thebibliography}{28}
\expandafter\ifx\csname
natexlab\endcsname\relax\def\natexlab#1{#1}\fi
\expandafter\ifx\csname bibnamefont\endcsname\relax
  \def\bibnamefont#1{#1}\fi
\expandafter\ifx\csname bibfnamefont\endcsname\relax
  \def\bibfnamefont#1{#1}\fi
\expandafter\ifx\csname citenamefont\endcsname\relax
\def\citenamefont#1{#1}\fi \expandafter\ifx\csname
url\endcsname\relax
  \def\url#1{\texttt{#1}}\fi
\expandafter\ifx\csname
urlprefix\endcsname\relax\def\urlprefix{URL }\fi
\providecommand{\bibinfo}[2]{#2}
\providecommand{\eprint}[2][]{\url{#2}}
\bibitem[{\citenamefont{Esaki and Tsu}(1970)}]{Esaki:IJRD:70:61}
\bibinfo{author}{\bibfnamefont{L.}~\bibnamefont{Esaki}} \bibnamefont{and}
  \bibinfo{author}{\bibfnamefont{R.}~\bibnamefont{Tsu}}, \bibinfo{journal}{IBM
  J. Res. Develop.} \textbf{\bibinfo{volume}{14}}, \bibinfo{pages}{61}
  (\bibinfo{year}{1970}).
\bibitem[{\citenamefont{Kazarinov and Suris}(1971)}]{Kazarinov:SPS:71:707}
\bibinfo{author}{\bibfnamefont{R.}~\bibnamefont{Kazarinov}} \bibnamefont{and}
  \bibinfo{author}{\bibfnamefont{R.}~\bibnamefont{Suris}},
  \bibinfo{journal}{Sov. Phys. Semicond.} \textbf{\bibinfo{volume}{5}},
  \bibinfo{pages}{707} (\bibinfo{year}{1971}).
\bibitem[{\citenamefont{Ktitorov et~al.}(1971)\citenamefont{Ktitorov, Simin,
  and Sindalovskii}}]{Ktitorov:FTT:71:2230}
\bibinfo{author}{\bibfnamefont{S.}~\bibnamefont{Ktitorov}},
  \bibinfo{author}{\bibfnamefont{G.}~\bibnamefont{Simin}}, \bibnamefont{and}
  \bibinfo{author}{\bibfnamefont{V.}~\bibnamefont{Sindalovskii}},
  \bibinfo{journal}{Fiz. tverd. Tela.} \textbf{\bibinfo{volume}{13}},
  \bibinfo{pages}{2230} (\bibinfo{year}{1971}).
\bibitem[{\citenamefont{Ignatov and Romanov}(1976)}]{Ignatov:PSSB:76:327}
\bibinfo{author}{\bibfnamefont{A.}~\bibnamefont{Ignatov}} \bibnamefont{and}
  \bibinfo{author}{\bibfnamefont{Y.}~\bibnamefont{Romanov}},
  \bibinfo{journal}{Phys. Stat. Sol. B} \textbf{\bibinfo{volume}{73}},
  \bibinfo{pages}{327} (\bibinfo{year}{1976}).
\bibitem[{\citenamefont{Faist et~al.}(1994)\citenamefont{Faist, Capasso, Sivco,
  Sirtori, Hutchinson, and Cho}}]{Faist:Sci:94:553}
\bibinfo{author}{\bibfnamefont{J.}~\bibnamefont{Faist}},
  \bibinfo{author}{\bibfnamefont{F.}~\bibnamefont{Capasso}},
  \bibinfo{author}{\bibfnamefont{D.}~\bibnamefont{Sivco}},
  \bibinfo{author}{\bibfnamefont{C.}~\bibnamefont{Sirtori}},
  \bibinfo{author}{\bibfnamefont{A.}~\bibnamefont{Hutchinson}},
  \bibnamefont{and} \bibinfo{author}{\bibfnamefont{A.}~\bibnamefont{Cho}},
  \bibinfo{journal}{Science} \textbf{\bibinfo{volume}{264}},
  \bibinfo{pages}{553} (\bibinfo{year}{1994}).
\bibitem[{\citenamefont{Beck et~al.}(2002)\citenamefont{Beck, Hofstetter,
  Aellen, Faist, Oesterle, Ilegems, Gini, and Melchior}}]{Beck:Sci:02:301}
\bibinfo{author}{\bibfnamefont{M.}~\bibnamefont{Beck}},
  \bibinfo{author}{\bibfnamefont{D.}~\bibnamefont{Hofstetter}},
  \bibinfo{author}{\bibfnamefont{T.}~\bibnamefont{Aellen}},
  \bibinfo{author}{\bibfnamefont{J.}~\bibnamefont{Faist}},
  \bibinfo{author}{\bibfnamefont{U.}~\bibnamefont{Oesterle}},
  \bibinfo{author}{\bibfnamefont{M.}~\bibnamefont{Ilegems}},
  \bibinfo{author}{\bibfnamefont{E.}~\bibnamefont{Gini}}, \bibnamefont{and}
  \bibinfo{author}{\bibfnamefont{H.}~\bibnamefont{Melchior}},
  \bibinfo{journal}{Science} \textbf{\bibinfo{volume}{295}},
  \bibinfo{pages}{301} (\bibinfo{year}{2002}).
\bibitem[{\citenamefont{K\"ohler et~al.}(2002)\citenamefont{K\"ohler,
  Tredicucci, Beltram, Beere, Davies, Linfield, Ritchie, Iotti, and
  Rossi}}]{Koehler:Nat:02:156}
\bibinfo{author}{\bibfnamefont{R.}~\bibnamefont{K\"ohler}},
  \bibinfo{author}{\bibfnamefont{A.}~\bibnamefont{Tredicucci}},
  \bibinfo{author}{\bibfnamefont{F.}~\bibnamefont{Beltram}},
  \bibinfo{author}{\bibfnamefont{H.}~\bibnamefont{Beere}},
  \bibinfo{author}{\bibfnamefont{G.}~\bibnamefont{Davies}},
  \bibinfo{author}{\bibfnamefont{E.}~\bibnamefont{Linfield}},
  \bibinfo{author}{\bibfnamefont{D.}~\bibnamefont{Ritchie}},
  \bibinfo{author}{\bibfnamefont{R.~C.} \bibnamefont{Iotti}}, \bibnamefont{and}
  \bibinfo{author}{\bibfnamefont{F.}~\bibnamefont{Rossi}},
  \bibinfo{journal}{Nature} \textbf{\bibinfo{volume}{417}},
  \bibinfo{pages}{156} (\bibinfo{year}{2002}).
\bibitem[{\citenamefont{Rochat et~al.}(2002)\citenamefont{Rochat, Ajili,
  Willenberg, Faist, Beere, Davies, Linfield, and
  Ritchie}}]{Rochat:APL:02:submitted}
\bibinfo{author}{\bibfnamefont{M.}~\bibnamefont{Rochat}},
  \bibinfo{author}{\bibfnamefont{L.}~\bibnamefont{Ajili}},
  \bibinfo{author}{\bibfnamefont{H.}~\bibnamefont{Willenberg}},
  \bibinfo{author}{\bibfnamefont{J.}~\bibnamefont{Faist}},
  \bibinfo{author}{\bibfnamefont{H.}~\bibnamefont{Beere}},
  \bibinfo{author}{\bibfnamefont{G.}~\bibnamefont{Davies}},
  \bibinfo{author}{\bibfnamefont{E.}~\bibnamefont{Linfield}}, \bibnamefont{and}
  \bibinfo{author}{\bibfnamefont{D.}~\bibnamefont{Ritchie}},
  \bibinfo{journal}{App. Phys. Lett.}  (\bibinfo{year}{2002}),
  \bibinfo{note}{unpublished}.
\bibitem[{\citenamefont{Sibille et~al.}(1990)\citenamefont{Sibille, Palmier,
  Wang, and Mollot}}]{Sibille:PRL:90:52}
\bibinfo{author}{\bibfnamefont{A.}~\bibnamefont{Sibille}},
  \bibinfo{author}{\bibfnamefont{J.}~\bibnamefont{Palmier}},
  \bibinfo{author}{\bibfnamefont{H.}~\bibnamefont{Wang}}, \bibnamefont{and}
  \bibinfo{author}{\bibfnamefont{F.}~\bibnamefont{Mollot}},
  \bibinfo{journal}{Phys. Rev. Lett.} \textbf{\bibinfo{volume}{64}},
  \bibinfo{pages}{52} (\bibinfo{year}{1990}).
\bibitem[{\citenamefont{Feldmann et~al.}(1992)\citenamefont{Feldmann, Leo,
  Shah, Miller, Cunningham, Meier, Plessen, Schulze, Thomas, and
  S.Schmitt-Rink}}]{Feldmann:PRB:92:7252}
\bibinfo{author}{\bibfnamefont{J.}~\bibnamefont{Feldmann}},
  \bibinfo{author}{\bibfnamefont{K.}~\bibnamefont{Leo}},
  \bibinfo{author}{\bibfnamefont{J.}~\bibnamefont{Shah}},
  \bibinfo{author}{\bibfnamefont{D.}~\bibnamefont{Miller}},
  \bibinfo{author}{\bibfnamefont{J.}~\bibnamefont{Cunningham}},
  \bibinfo{author}{\bibfnamefont{T.}~\bibnamefont{Meier}},
  \bibinfo{author}{\bibfnamefont{G.}~\bibnamefont{Plessen}},
  \bibinfo{author}{\bibfnamefont{A.}~\bibnamefont{Schulze}},
  \bibinfo{author}{\bibfnamefont{P.}~\bibnamefont{Thomas}}, \bibnamefont{and}
  \bibinfo{author}{\bibnamefont{S.Schmitt-Rink}}, \bibinfo{journal}{Phys. Rev.
  B} \textbf{\bibinfo{volume}{46}}, \bibinfo{pages}{7252}
  (\bibinfo{year}{1992}).
\bibitem[{\citenamefont{Keay et~al.}(1995)\citenamefont{Keay, Zeuner, Allen,
  Maranowski, Gossard, Bhattacharya, and Rodwell}}]{Keay:PRL:95:4102}
  \bibinfo{author}{\bibfnamefont{B.}~\bibnamefont{Keay}},
  \bibinfo{author}{\bibfnamefont{S.}~\bibnamefont{Zeuner}},
  \bibinfo{author}{\bibfnamefont{S.}~\bibnamefont{Allen}},
  \bibinfo{author}{\bibfnamefont{K.}~\bibnamefont{Maranowski}},
  \bibinfo{author}{\bibfnamefont{A.}~\bibnamefont{Gossard}},
  \bibinfo{author}{\bibfnamefont{U.}~\bibnamefont{Bhattacharya}},
  \bibnamefont{and} \bibinfo{author}{\bibfnamefont{M.}~\bibnamefont{Rodwell}},
  \bibinfo{journal}{Phys. Rev. Lett.} \textbf{\bibinfo{volume}{75}},
  \bibinfo{pages}{4102} (\bibinfo{year}{1995}).
\bibitem[{\citenamefont{Unterrainer et~al.}(1996)\citenamefont{Unterrainer,
  Keay, Wanke, Allen, Leonard, G.~Medeiros-Ribeiro, and
  Rodwell}}]{Unterrainer:PRL:96:2973}
  \bibinfo{author}{\bibfnamefont{K.}~\bibnamefont{Unterrainer}},
  \bibinfo{author}{\bibfnamefont{B.}~\bibnamefont{Keay}},
  \bibinfo{author}{\bibfnamefont{M.}~\bibnamefont{Wanke}},
  \bibinfo{author}{\bibfnamefont{S.}~\bibnamefont{Allen}},
  \bibinfo{author}{\bibfnamefont{D.}~\bibnamefont{Leonard}},
  \bibinfo{author}{\bibfnamefont{U.~B.} \bibnamefont{G.~Medeiros-Ribeiro}},
  \bibnamefont{and} \bibinfo{author}{\bibfnamefont{M.}~\bibnamefont{Rodwell}},
  \bibinfo{journal}{Phys. Rev. Lett.} \textbf{\bibinfo{volume}{76}},
  \bibinfo{pages}{2973} (\bibinfo{year}{1996}).
\bibitem[{\citenamefont{Kazarinov and Suris}(1972)}]{Kazarinov:SPS:72:120}
\bibinfo{author}{\bibfnamefont{R.}~\bibnamefont{Kazarinov}} \bibnamefont{and}
  \bibinfo{author}{\bibfnamefont{R.}~\bibnamefont{Suris}},
  \bibinfo{journal}{Sov. Phys. Semicond.} \textbf{\bibinfo{volume}{6}},
  \bibinfo{pages}{120} (\bibinfo{year}{1972}).
\bibitem[{\citenamefont{Kohn}(1959)}]{Kohn:PRL:59:809}
\bibinfo{author}{\bibfnamefont{W.}~\bibnamefont{Kohn}}, \bibinfo{journal}{Phys.
  Rev.} \textbf{\bibinfo{volume}{15}}, \bibinfo{pages}{809}
  (\bibinfo{year}{1959}).
\bibitem[{\citenamefont{Kohn and Luttinger}(1957)}]{Kohn:PRL:57:590}
\bibinfo{author}{\bibfnamefont{W.}~\bibnamefont{Kohn}} \bibnamefont{and}
  \bibinfo{author}{\bibfnamefont{J.~M.} \bibnamefont{Luttinger}},
  \bibinfo{journal}{Phys. Rev.} \textbf{\bibinfo{volume}{108}},
  \bibinfo{pages}{590} (\bibinfo{year}{1957}).
\bibitem[{\citenamefont{Wacker}(1998)}]{Wacker:ttsn:98:321}
\bibinfo{author}{\bibfnamefont{A.}~\bibnamefont{Wacker}}, in
  \emph{\bibinfo{booktitle}{Theory of Transport Properties of Semiconductor
  Nanostructures}}, edited by
  \bibinfo{editor}{\bibfnamefont{E.}~\bibnamefont{Sch\"oll}}
  (\bibinfo{publisher}{Chapman and Hall, London}, \bibinfo{year}{1998}), p.
  \bibinfo{pages}{321}.
\bibitem[{\citenamefont{Wacker et~al.}(1997)\citenamefont{Wacker, Jauho,
  Zeuner, and Allen}}]{Wacker:PRB:97:13268}
\bibinfo{author}{\bibfnamefont{A.}~\bibnamefont{Wacker}},
  \bibinfo{author}{\bibfnamefont{A.-P.} \bibnamefont{Jauho}},
  \bibinfo{author}{\bibfnamefont{S.}~\bibnamefont{Zeuner}}, \bibnamefont{and}
  \bibinfo{author}{\bibfnamefont{S.~J.} \bibnamefont{Allen}},
  \bibinfo{journal}{Phys. Rev. B} \textbf{\bibinfo{volume}{97}},
  \bibinfo{pages}{13268} (\bibinfo{year}{1997}).
\bibitem[{\citenamefont{Unterrainer}(2000)}]{Unterrainer:ssv66:00:4}
\bibinfo{author}{\bibfnamefont{K.}~\bibnamefont{Unterrainer}}, in
  \emph{\bibinfo{booktitle}{Intersubband transitions in quantum wells: Physics
  and device applications II}}, edited by
  \bibinfo{editor}{\bibfnamefont{H.}~\bibnamefont{Liu}} \bibnamefont{and}
  \bibinfo{editor}{\bibfnamefont{F.}~\bibnamefont{Capasso}}
  (\bibinfo{publisher}{Academic Press}, \bibinfo{year}{2000}),
  vol.~\bibinfo{volume}{66}, chap.~\bibinfo{chapter}{4}, p.
  \bibinfo{pages}{139}.
\bibitem[{\citenamefont{Schmidt et~al.}()\citenamefont{Schmidt, Willenberg,
  Faist, and D\"ohler}}]{Schmidt:unpublished}
\bibinfo{author}{\bibfnamefont{A.~B.} \bibnamefont{Schmidt}},
  \bibinfo{author}{\bibfnamefont{H.}~\bibnamefont{Willenberg}},
  \bibinfo{author}{\bibfnamefont{J.}~\bibnamefont{Faist}}, \bibnamefont{and}
  \bibinfo{author}{\bibfnamefont{G.~H.} \bibnamefont{D\"ohler}},
  \bibinfo{note}{unpublished}.
\bibitem[{\citenamefont{Rott et~al.}(1999)\citenamefont{Rott, Binder, Linder,
  and D\"ohler}}]{Rott:PRB:99:7334}
\bibinfo{author}{\bibfnamefont{S.}~\bibnamefont{Rott}},
  \bibinfo{author}{\bibfnamefont{P.}~\bibnamefont{Binder}},
  \bibinfo{author}{\bibfnamefont{N.}~\bibnamefont{Linder}}, \bibnamefont{and}
  \bibinfo{author}{\bibfnamefont{G.~H.} \bibnamefont{D\"ohler}},
  \bibinfo{journal}{Phys. Rev. B} \textbf{\bibinfo{volume}{59}},
  \bibinfo{pages}{7334} (\bibinfo{year}{1999}).
\bibitem[{Wil({\natexlab{a}})}]{Willenberg:montecarlo}
\bibinfo{note}{Provided momentum and energy relaxation times are deduced from
  microscopic scattering rates, equation (\ref{ignatov}) is in excellent
  agreement with Monte-Carlo calculations, whereas the single relaxation time
  approximation systematically overestimates the semi-classical gain,
  unpublished results}.
\bibitem[{\citenamefont{Metzner et~al.}(1995)\citenamefont{Metzner, Schr\"ufer,
  Wieser, Luber, Kneissl, and D\"ohler}}]{Metzner:PRB:95:5106}
\bibinfo{author}{\bibfnamefont{C.}~\bibnamefont{Metzner}},
  \bibinfo{author}{\bibfnamefont{K.}~\bibnamefont{Schr\"ufer}},
  \bibinfo{author}{\bibfnamefont{U.}~\bibnamefont{Wieser}},
  \bibinfo{author}{\bibfnamefont{M.}~\bibnamefont{Luber}},
  \bibinfo{author}{\bibfnamefont{M.}~\bibnamefont{Kneissl}}, \bibnamefont{and}
  \bibinfo{author}{\bibfnamefont{G.~H.} \bibnamefont{D\"ohler}},
  \bibinfo{journal}{Phys. Rev. B} \textbf{\bibinfo{volume}{51}},
  \bibinfo{pages}{5106} (\bibinfo{year}{1995}).
\bibitem[{\citenamefont{Willenberg et~al.}(2002)\citenamefont{Willenberg,
  Wolst, Elpelt, Geisselbrecht, Malzer, and
  D\"ohler}}]{Willenberg:PRB:02:35328}
\bibinfo{author}{\bibfnamefont{H.}~\bibnamefont{Willenberg}},
  \bibinfo{author}{\bibfnamefont{O.}~\bibnamefont{Wolst}},
  \bibinfo{author}{\bibfnamefont{R.}~\bibnamefont{Elpelt}},
  \bibinfo{author}{\bibfnamefont{W.}~\bibnamefont{Geisselbrecht}},
  \bibinfo{author}{\bibfnamefont{S.}~\bibnamefont{Malzer}}, \bibnamefont{and}
  \bibinfo{author}{\bibfnamefont{G.~H.} \bibnamefont{D\"ohler}},
  \bibinfo{journal}{Phys. Rev. B} \textbf{\bibinfo{volume}{65}},
  \bibinfo{pages}{35328} (\bibinfo{year}{2002}).
\bibitem[{Wil({\natexlab{b}})}]{Willenberg:disorder}
\bibinfo{note}{Assume $\Delta n=0$. A pair of subbands with energy separation
  $\Delta\epsilon_{ij}<\epsilon$ resembles subbands with a population inversion
  $\Delta n/n>0$, as the quasi-Fermi-level in the upper subband is higher,
  regardless of its absolute position, than in the lower state, which follows
  from $\Delta\epsilon_{ij}<\Phi_{i}-\Phi_{j}$. However, subbands spaced by
  more than the average transition energy, $\Delta\epsilon_{ij}>\epsilon$,
  contribute with a line shape corresponding to $\Delta n/n<0$. Hence,
  inhomogeneous broadening does not average over the dispersive profile.}
\bibitem[{\citenamefont{Campman et~al.}(1996)\citenamefont{Campman, Schmidt,
  Imamoglu, and Gossard}}]{Campman:APL:96:2554}
\bibinfo{author}{\bibfnamefont{K.}~\bibnamefont{Campman}},
  \bibinfo{author}{\bibfnamefont{H.}~\bibnamefont{Schmidt}},
  \bibinfo{author}{\bibfnamefont{A.}~\bibnamefont{Imamoglu}}, \bibnamefont{and}
  \bibinfo{author}{\bibfnamefont{A.}~\bibnamefont{Gossard}},
  \bibinfo{journal}{Appl. Phys. Lett.} \textbf{\bibinfo{volume}{69}},
  \bibinfo{pages}{2554} (\bibinfo{year}{1996}).
\bibitem[{\citenamefont{Warburton et~al.}(1996)\citenamefont{Warburton, Gauer,
  Wixforth, Kotthaus, Brar, and Kroemer}}]{Warburton:SM:96:365}
\bibinfo{author}{\bibfnamefont{R.}~\bibnamefont{Warburton}},
  \bibinfo{author}{\bibfnamefont{C.}~\bibnamefont{Gauer}},
  \bibinfo{author}{\bibfnamefont{A.}~\bibnamefont{Wixforth}},
  \bibinfo{author}{\bibfnamefont{J.}~\bibnamefont{Kotthaus}},
  \bibinfo{author}{\bibfnamefont{B.}~\bibnamefont{Brar}}, \bibnamefont{and}
  \bibinfo{author}{\bibfnamefont{H.}~\bibnamefont{Kroemer}},
  \bibinfo{journal}{Superlatt.Microstruct.} \textbf{\bibinfo{volume}{19}},
  \bibinfo{pages}{365} (\bibinfo{year}{1996}).
\bibitem[{\citenamefont{Luin et~al.}(2001)\citenamefont{Luin, Pellegrini,
  Beltram, Marcadet, and Sirtori}}]{Luin:PRB:01:041306-1}
  \bibinfo{author}{\bibfnamefont{S.}~\bibnamefont{Luin}},
  \bibinfo{author}{\bibfnamefont{V.}~\bibnamefont{Pellegrini}},
  \bibinfo{author}{\bibfnamefont{F.}~\bibnamefont{Beltram}},
  \bibinfo{author}{\bibfnamefont{X.}~\bibnamefont{Marcadet}}, \bibnamefont{and}
  \bibinfo{author}{\bibfnamefont{C.}~\bibnamefont{Sirtori}},
  \bibinfo{journal}{Phys. Rev. B} \textbf{\bibinfo{volume}{64}},
  \bibinfo{pages}{041306} (\bibinfo{year}{2001}).
\bibitem[{\citenamefont{Kroemer}()}]{Kroemer:CondMat:00:xx}
\bibinfo{author}{\bibfnamefont{H.}~\bibnamefont{Kroemer}},
  \bibinfo{note}{\emph{On the nature of the negative-conductivity resonance in
  a superlattice Bloch oscillator}, cond-mat, 0007482, (2001)}.
\end{thebibliography}
\begin{widetext}
\end{widetext}
\section*{Appendix: Mathematical Details}

\subsection{Non-diagonal part $f^{ij}_{kk'}$}\label{appendixa}

Equation (\ref{26f}) governs the dynamics of the non-diagonal part
in $k,k'$
\begin{eqnarray} i\hbar s f^{ij}_{kk'}&&\approx i\hbar s
\rho^{ij}_{kk'}(0)+ \sum_{m}(V_{kk'}^{im} f^{mj}_{k'}-
f^{im}_{k}V_{kk'}^{mj})\nonumber\\&&+ \sum_{m}(H_{k}^{im}
f^{mj}_{kk'}- f^{im}_{kk'}H_{k'}^{mj})\label{nd}\,.
\end{eqnarray}
Assuming intra-well scattering only, the second and third term on
the RHS yield
\begin{eqnarray*}
\sum_{m}&&(V_{kk'}^{im} f^{mj}_{k'}- f^{im}_{k}V_{kk'}^{mj})
\approx  V_{kk'}^{ii} f^{ij}_{k'}- f^{ij}_{k}V_{kk'}^{jj} \\
\sum_{m}&&(H_{k}^{im} f^{mj}_{kk'}-
f^{im}_{kk'}H_{k'}^{mj})=\nonumber\\&&= (\epsilon_{ik}
-\epsilon_{jk'}) f^{ij}_{kk'} +
\hbar\Omega_{ij}(f^{jj}_{kk'}-f^{ii}_{kk'})\,.
\end{eqnarray*}
Neglecting the non-diagonal matrix element $\rho^{ij}_{kk'}(0)$
and taking the Laplace average equation (\ref{nd}) gives
\begin{eqnarray}
f^{ij}_{kk'}&&=-\left(\mathcal{P}\frac{1}{\epsilon_{ik}
-\epsilon_{jk'}}-i\pi\delta(\epsilon_{ik}-\epsilon_{jk'})\right)
\\
&&\times\left(\hbar\Omega_{ij}(f^{jj}_{kk'}-f^{ii}_{kk'})+
V_{kk'}^{ii} f^{ij}_{k'}-
f^{ij}_{k}V_{kk'}^{jj})\right)\,.\nonumber
\end{eqnarray}
The non-diagonal $f^{ij}_{kk'}$ still depends on
$f_{kk'}^{jj}-f_{kk'}^{ii}$. The coherences between states $k$ and
$k'$ within the same subband are derived from the special version
for $i=j$ of equation~(\ref{nd})
\begin{eqnarray}
i\hbar s f^{ii}_{kk'}&&\approx i\hbar s \rho^{ii}_{kk'}(0)+
\sum_{m}(V_{kk'}^{im} f^{mi}_{k'}- f^{im}_{k}V_{kk'}^{mi})
\nonumber\\&&+\sum_{m}(H_{k}^{im} f^{mi}_{kk'}-
f^{im}_{kk'}H_{k'}^{mi})\,,
\end{eqnarray}
where the second and third term are evaluated as
\begin{eqnarray*}
\sum_{m}&&(V_{kk'}^{im} f^{mi}_{k'}- f^{im}_{k}V_{kk'}^{mi})=
V_{kk'}^{ii} (f^{ii}_{k'}- f^{ii}_{k})\\ \sum_{m}&&(H_{k}^{im}
f^{mi}_{kk'}- f^{im}_{kk'}H_{k'}^{mi}) =\\&&=(\epsilon_{ik}
-\epsilon_{ik'}) f^{ii}_{kk'}+
\hbar\Omega_{ij}f^{ji}_{kk'}-f^{ij}_{kk'}\hbar\Omega_{ji}\\&&
\approx (\epsilon_{ik} -\epsilon_{ik'}) f^{ii}_{kk'}
\end{eqnarray*}
and terms of higher order in the tunneling matrix element
corresponding to multiple tunneling processes are neglected.
Taking the Laplace average yields
\begin{eqnarray} f^{ii}_{kk'} &&= -\mathcal{P}\frac{1}{\epsilon_{ik}
-\epsilon_{ik'}}V_{kk'}^{ii}(f^{ii}_{k'}- f^{ii}_{k})+
\\&&+i\pi\delta(\epsilon_{ik}
-\epsilon_{ik'})V_{kk'}^{ii}(f^{ii}_{k'}- f^{ii}_{k})\approx
\frac{V_{kk'}^{ii}(f^{ii}_{k}- f^{ii}_{k'})}{\epsilon_{ik}
-\epsilon_{ik'}}\,.\nonumber
\end{eqnarray}
The last term vanishes as either the $\delta$-function or the
difference in populations is zero. Placing the approximations for
$f^{ii}_{kk'}$ and $f^{jj}_{kk'}$ in equation~(\ref{nd}) and
neglecting the principal value yields
\begin{widetext}
\begin{equation} f^{ij}_{kk'} = i\pi\delta(\epsilon_{ik}
-\epsilon_{jk'})\left(\hbar\Omega_{ij}\left(\frac{V_{kk'}^{jj}(f^{jj}_{k}-
f^{jj}_{k'})}{\epsilon_{jk} -\epsilon_{jk'}}-
\frac{V_{kk'}^{ii}(f^{ii}_{k}- f^{ii}_{k'})}{\epsilon_{ik}
-\epsilon_{ik'}}\right)+ V_{kk'}^{ii} f^{ij}_{k'}-
f^{ij}_{k}V_{kk'}^{jj}\right)\,.
\end{equation}
\end{widetext}

\subsection{Diagonal part $f^{ij}_{k}$}\label{appendixb}

Equation (\ref{25f}) determines the dynamics of the diagonal part
in $k$
\begin{eqnarray}
i\hbar s f^{ij}_{k}&& = i\hbar s\rho^{ij}_{k}(0) +
\sum_{m}(H_{k}^{im} f^{mj}_{k}-f^{im}_{k}H_{k}^{mj})\nonumber\\&&+
\sum_{m,k'}(V_{kk'}^{im}\left(f\right)_{k'k}^{mj}
-\left(f\right)_{kk'}^{im} V_{k'k}^{mj})\label{25z}\,,
\end{eqnarray}
where $(f)$ denotes the previous approximations of the
non-diagonal part. The second term on the RHS is given by
\begin{eqnarray*}
\sum_{m}&&(H_{k}^{im} f^{mj}_{k}-
f^{im}_{k}H_{k}^{mj})=\nonumber\\&&=(\epsilon_{ik} -\epsilon_{jk})
f^{ij}_{k}+\hbar\Omega_{ij}(f_{k}^{jj} - f_{k}^{ii})
\end{eqnarray*}
Performing the Laplace limit $s\rightarrow 0$ we obtain
\begin{eqnarray}
(\epsilon_{ik} -\epsilon_{jk}) f^{ij}_{k}&& =
\hbar\Omega_{ij}(f_{k}^{ii} - f_{k}^{jj})\\&&-
\sum_{k'}(V_{kk'}^{ii}\left(f\right)_{k'k}^{ij}
-\left(f\right)_{kk'}^{ij} V_{k'k}^{jj})\nonumber\label{almost}
\end{eqnarray}
If one assumes no correlation between scattering events due to the
impurities in different wells, \emph{i.e.} were to drop terms
containing the product $V_{kk'}^{ii}V_{kk'}^{jj}$ for $i\neq j$,
the product of scattering potentials and the approximated
non-diagonal part becomes
\begin{eqnarray*}
V_{kk'}^{ii}\left(f\right)_{k'k}^{ij}=&&i\pi
\delta(\epsilon_{ik'}-\epsilon_{jk})|V_{kk'}^{ii}|^2\\ &&\times
(f^{ij}_{k}+\frac{\hbar\Omega_{ij}(f^{ii}_{k'}-f^{ii}_{k})}{\epsilon_{ik'}
-\epsilon_{ik}})
\end{eqnarray*}
and similarly
\begin{eqnarray*}
\left(f\right)_{kk'}^{ij}V_{k'k}^{jj}=
&&i\pi\delta(\epsilon_{ik}-\epsilon_{jk'})|V_{kk'}^{jj}|^2\\
&&\times
(f^{ij}_{k}+\frac{\hbar\Omega_{ij}(f^{jj}_{k}-f^{jj}_{k'})}{\epsilon_{jk}-\epsilon_{jk'}})\nonumber
\label{}
\end{eqnarray*}
Rewriting equation~(\ref{almost}) and sorting terms finally leads
to an equation for the relevant coherences between the two states,
that determine transport properties such as the current density
\begin{widetext}
\begin{eqnarray} (\epsilon_{ik} -\epsilon_{jk}) f^{ij}_{k} -
&& i\pi f^{ij}_{k} \overbrace{\sum_{k'}
\delta(\epsilon_{ik'}-\epsilon_{jk})|V_{kk'}^{ii}|^2
+\delta(\epsilon_{ik}-\epsilon_{jk'})|V_{kk'}^{jj}|^2}^{\mbox{transition
broadening}} = \overbrace{\hbar\Omega_{ij}(f_{k}^{ii} -
f_{k}^{jj})}^{\mbox{population inversion}\rightarrow
f^{qc}}\label{central}
\\&&+\underbrace{i\pi\sum_{k'}
\delta(\epsilon_{ik'}-\epsilon_{jk})|V_{kk'}^{ii}|^2
\frac{\hbar\Omega_{ij}}{\epsilon_{ik'}-\epsilon_{ik}}
(f^{ii}_{k'}-f^{ii}_{k})
+\delta(\epsilon_{ik}-\epsilon_{jk'})|V_{kk'}^{jj}|^2
\frac{\hbar\Omega_{ij}}{\epsilon_{jk}-\epsilon_{jk'}}
(f^{jj}_{k}-f^{jj}_{k'})}_{\mbox{Bloch type
contribution}\rightarrow f^{bo}}\nonumber\,.
\end{eqnarray}
\end{widetext}
It corresponds to the results of Kazarinov and Suris
\cite{Kazarinov:SPS:72:120} after performing a non-correlated
ensemble average on their equation and specifying on a two-level
system.

\subsection{Absorption}

The ac-field induced coherence is given by the transformation
\begin{equation} \delta f_{k,\omega}^{21} = -
\frac{edf_\omega}{\hbar\omega}f^{21}_{k,\epsilon-\hbar\omega}\,,
\end{equation}
which yields
\begin{eqnarray*} \delta f_k^{21} = &&-
\frac{edf_\omega}{\hbar\omega}\Bigl\{\frac{\hbar\Omega_{21}(f^{22}_k-f^{11}_k)}{\epsilon-\hbar\omega
-i(\gamma_{k}^{2}+\gamma_{k}^{1})}
\\ &&+
\frac{i\hbar\Omega_{21}(\gamma_{k}^{2}(f^{22}_{k-}-f^{22}_k)
-\gamma_{k}^{1}(f^{11}_{k+}-f^{11}_k))}{(\epsilon-\hbar\omega)(\epsilon-\hbar\omega
-i(\gamma_{k}^{2}+\gamma_{k}^{1}))}\Bigr\}\,.
\end{eqnarray*}
As each $\delta f =\delta(f^{\rm qc}+f^{\rm bo})$ the absorption
consists of two contributions
\begin{equation}
\alpha(\omega) = \alpha^{qc}(\omega)+\alpha^{bo}(\omega)\,,
\end{equation}
like the two contributions on the RHS of equation (\ref{central})
add up in respect to the current density. We obtain
\begin{widetext}
\begin{eqnarray}
\alpha^{qc}(\omega) =\frac{e^2 d^2
|\Omega_{21}|^2}{\varepsilon_0n_rc\omega}\sum_k
\left(\frac{\hbar\tau_k^{-1}}
{(\epsilon+\hbar\omega)^2+(\hbar\tau_k^{-1})^2}
\underbrace{-\frac{\hbar\tau_k^{-1}}
{(\epsilon-\hbar\omega)^2+(\hbar\tau_k^{-1})^2}}_{\text{main
contribution}}\right) (f_{k}^{22}-f_{k}^{11})
\end{eqnarray}
and
\begin{eqnarray} \alpha^{bo}(\omega) =\frac{e^2 d^2
|\Omega_{21}|^2}{\varepsilon_0n_rc\omega}\sum_k
\left(\frac{\gamma_{k}^2 (f^{22}_{l-}-f^{22}_k)
-\gamma_{k}^1(f^{11}_{l+}-f^{11}_{k})}
{(\epsilon+\hbar\omega)^2+(\hbar\tau_k^{-1})^2} -
\underbrace{\frac{\gamma_{k}^2(f^{22}_{k-}-f^{22}_k)
-\gamma_{k}^1(f^{11}_{k+}-f^{11}_{k})}
{(\epsilon-\hbar\omega)^2+(\hbar\tau_k^{-1})^2}}_{\text{main
contribution}}\right)\,,
\end{eqnarray}
\end{widetext}
where we set
$k_\pm=\hbar^{-1}\sqrt{2m^*(\epsilon_k\pm(\epsilon-\hbar\omega))}$,
$l_\pm=\hbar^{-1}\sqrt{2m^*(\epsilon_k\pm(\epsilon+\hbar\omega))}$
and $\hbar\tau_k^{-1} = \gamma_k^1+\gamma_k^2$. The first term,
$\alpha^{qc}(\omega)$, depends on the difference in population of
equivalent $k$-states and yields the usual Lorentzian line shape
for the gain profile in case of a population inversion. The second
contribution, $\alpha^{bo}(\omega)$, contains differences of
populations between different $k$-states within the respective
subband state and will be discussed later.

In the following, we will omit the non-resonant contribution. The
contribution accounts for the situation where the upper state lies
below the lower state. However, it is not neglected in the
numerical calculations as it is important to prevent the
divergence at $\omega=0$, which alters the line-shape for
$\hbar\omega \sim \mathcal{O}(\gamma)$, \emph{i.e.} in the
far-infrared or terahertz regime. If one regards the resonant
contribution only, the inversion gain reads
\begin{eqnarray*}
\alpha^{qc}(\omega) =\frac{e^2 d^2
|\Omega_{21}|^2}{\varepsilon_0n_rc\omega}\sum_k
\frac{\hbar\tau_k^{-1}(f_{k}^{11}-f_{k}^{22})}
{(\epsilon-\hbar\omega)^2+(\hbar\tau_k^{-1})^2}
\end{eqnarray*}
whereas the Bloch type contribution gives
\begin{eqnarray*}
\alpha^{bo}(\omega)=\frac{e^2 d^2
|\Omega_{21}|^2}{\varepsilon_0n_rc\omega}\sum_k
\frac{\gamma_{k}^1(f^{11}_{k+}-f^{11}_{k})
-\gamma_{k}^2(f^{22}_{k-}-f^{22}_{k})}
{(\epsilon-\hbar\omega)^2+(\hbar\tau_k^{-1})^2}\nonumber\,.
\end{eqnarray*}
The difficulty in assigning a path of transitions of the electron
to the latter expression for $\alpha^{bo}(\omega)$ is resolved by
adding both contributions (\emph{cf.} equation (\ref{diag})).

\end{document}